# Thinging-Oriented Modeling of Unmanned Aerial Vehicles

Sabah Al-Fedaghi[1], Jassim Al-Fadhli[2]
Computer Engineering Department
Kuwait University
Kuwait

*Abstract*—In recent years, there has been a dramatic increase in both practical and research applications of unmanned aerial vehicles (UAVs). According to the literature, there is a need in this area to develop a more refined model of UAV system architecture—in other words, a conceptual model that defines the system's structure and behavior. The existing models mostly are fractional and do not account for the entire important dynamic attributes. Progress in this area could reduce ambiguity and increase reliability in the design of such systems. This paper aims to advance the modeling of UAV system architecture by adopting a conceptual model called a thinging (abstract) machine in which all of the UAV's software and hardware components are viewed in terms of the flow of things and five generic operations. We apply this model to a real case study of a drone. The results—an integrated conceptual representation of the drone—support the viability of this approach.

## I. INTRODUCTION

Individuals are becoming more reliant on automated systems for a wide range of applications. According to Culus Schellekens, and Smeets [1], "It looks like our 21st century will be the century of robots, with a lot of buzz concerning a fast-growing subfamily of these machines, namely drones." The recent drone attacks on Saudi Arabia's oil installations highlight the importance of giving this technology high research priority. The introduction of unmanned aerial vehicles (UAVs) has raised profound questions around the world with regard to issues such as accountability, transparency, privacy, legality, use of force, and safety [2]. In recent years, although UAVs have been an in-demand research topic, "there still exist many unanswered questions" [3].

From a technical standpoint, unmanned airplanes can be categorized as UAVs, remotely piloted vehicles, or drones. These types differ mainly in the type of the mission, the size, and (importantly) the level of autonomy in their operation [4]. The term "drone," which is in general use among both the media and the public, refers to all types of UAVs.

Nowadays, the number of possible uses of UAVs is large and increasing [5]. One planned development in this direction is the use of a centimeter-scale quadcopter with a driving application over vast regions. An immense number of such vehicles are used for various purposes (e.g., providing climatic and meteorological data) [3]. One especially important application of UAVs is carrying out so-called D missions [6]: those that are dangerous, dirty, or dull.

This paper focuses on the high-level modeling and control of UAVs. Its general objective is to offer a schematic language for UAVs so as to enhance the understanding of their functionalities. The understanding of technology is a constitutive part of human life and helps to address issues of survival and improve the use and practice of technology. According to Sellars [7], contemporary society hangs together largely through technology. This context requires understanding "both the practice of designing and creating artifacts (in a wide sense, including artificial processes and systems) and the nature of the things so created" [8]. Accordingly, there is a need to develop a language that ensures good technical specification. Such a specification is similar to a script for performing a task in that it allows stakeholders (engineers, team members, legislators, technocrats, managers, officials, etc.) to understand the roles they need to play and that helps them to avoid either stepping on one another's toes or overlooking a critical piece of information [9].

This paper's specific objective is thus to provide a modeling language that can be used to specify the system architecture as an integral phase of the UAV development process. The system architecture is the conceptual model that defines the structural, behavioral, and other views of a system [10]. Indeed, it is necessary to develop, for UAVs, both "architecture generation and assessment models. Architecture assessment models that presently exist tend to be fractional and do not account for all dynamic attributes that should be considered in the architecture assessment" [11]. Further development in this area can reduces systems' ambiguity and increases their tangibility.

To accomplish these aims, we propose applying a diagrammatic modeling technique called a thinging machine (TM). This modeling apparatus is viable in the area of UAV systems architecture because it can provide a precise description of the total system. This claim is substantiated by contrasting TM models with the current UAV diagramming methods. Additionally, this paper shows the feasibility of the TM approach by describing a real case study.

Section II includes a partial survey of the works in the area of diagram-based modeling for UAVs. Section III provides a review of the TM modeling tool, and a detailed example is given in Section IV. Section V describes the case study, in which an actual drone is modeled. Section VI clarify the TM model itself and Section VII present the application of TM in the case study and describe the possible utilization of the resulting model in a simulation. Section VIII discuss the simulation aspects, and finally Section IX is the conclusion.

## II. SOME RELATED WORKS

Extensive research has been conducted on the conceptual modeling of UAVs at all levels [12]. Drones provide unprecedented levels of access to airspace, and such new access could fundamentally change business, shipment, and travel [13]. However, we focus here on a few architectural models that facilitate contrasts with the TM approach. A comprehensive survey of the field can be found in Renaul [11].

Pastor et al. [6] presented a hardware/software architecture for UAVs, using block diagrams to provide a general view of the architecture of a mission-control computer and ad hoc diagrams to describe operational scenarios. This is an example of such a scenario: The mission control decides to take a georeferenced video. For this task, it will need the services provided by storage, the flight-computer system, and the camera and sensing modules.

Diem, Hien, and Khanh [14] sought to analyze, design, and implement controllers of a standard UAV platform; they thus adopted a model-driven architecture with a real-time unified modeling language (UML). This architecture contains three models that are used to separate the specifications for a system's operation: the computation-independent, platform-independent, and platform-specific models. To capture the general requirements based on the object-oriented paradigm, Diem, Hien, and Khanh [14] presented a model with abstract classes using UML stereotypes and a class diagram [15] in order to describe the main functional components of quadrotor UAVs (see Fig. 1-3). According to Diem, Hien, and Khanh [14], in Fig. 1, "The Guidance System's class is responsible for giving the desired trajectory for a quadrotor UAV to follow. This responsibility is completed by taking the desired waypoints defining pre-mission with the possible inclusion of external environmental disturbances issued from the Air Environment's class; then, it generates a path for this quadrotor UAV to follow.

This is the level of modeling and description that TM modeling targets. The style of diagramming in Fig. 1-3 is included to provide a contrast with the TM diagrams for the model UAV, as developed later in this paper.

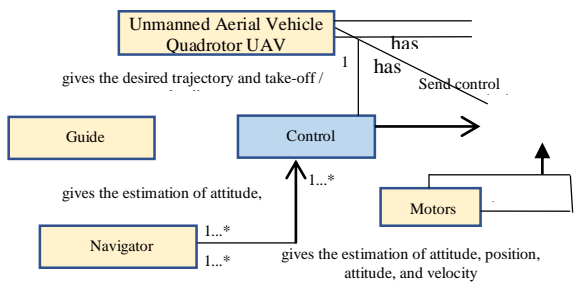

Fig. 1. A UML Class Diagram for Presenting the main Functional Components of Quadrotor UAVs (Redrawn and Adapted from [14]).

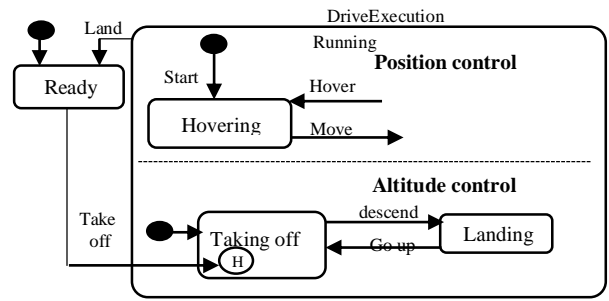

Fig. 2. Local State Machine for the Drive use case. (Redrawn and Adapted from [14]).

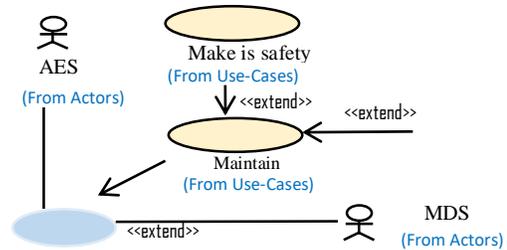

Fig. 3. Main use-case Model for a Quadrotor UAV. (Redrawn and Adapted from [14]).

Although nothing prevents a diagram from presenting multiple views of a system, these diagrams are heterogeneous and contain awkward symbols; thus, there is a need for more systematic depictions that, according to [16], help to meet the challenge of defining a single coherent architecture. TM both presents the totality of the system in a conceptual form and distinguishes between a model's static and dynamic aspects.

## III. THINGING MACHINES

We adopt a conceptual model that is centered on a system's things and (abstract) machines. The philosophical foundation of this approach is Heidegger's notion of thinging [17]. According to Riemer, Johnston, Hovorka, and Indulska [18], Heidegger's philosophy gives an alternative analysis of "(1) eliciting knowledge of routine activities, (2) capturing knowledge from domain experts and (3) representing organizational reality in authentic ways" [18]. More information about TM's philosophical foundation can be found in Al-Fedaghi [19–21].

The simplest type of the thing/machine combination is a TM, which is a generalization of the known input-process-output model. In a TM, the flow of things is the exclusive conceptual movement among the five operations (stages), as shown in Fig. 4. A thing is created, processed, released, transferred, and/or received in a machine.

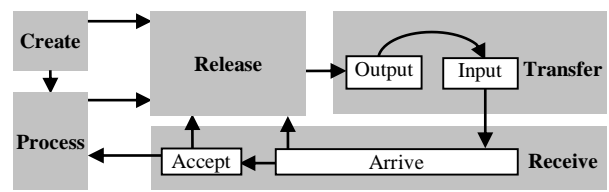

Fig. 4. A Thinging Machine.

Accordingly, the stages of a TM can be described as operations that transform, modify, or otherwise affect things abstractly or concretely. They are as follows.

- **Arrive:** A thing flows to a new machine (e.g., packets arrive at a router's port).

- **Accept**: A thing enters a TM after arrival; on the assumption that all arriving things are accepted, arrive and accept can be combined as the **Receiving** stage.

- **Release**: A thing is marked as ready for transfer outside the machine (e.g., in an airport, passengers wait to board after passport clearance).

- **Process**: A thing's descriptions are changed (rather than the thing itself).

- **Create**: A new thing is created (e.g., a forward packet is generated in a machine).

- **Transfer**: A thing is input to or output from a machine.

TM also includes *triggering* (denoted by a dashed arrow), or the initiation of a new flow (e.g., electricity triggers a flow of air). TM modeling is used in many applications (e.g., see Al-Fedaghi [22–25]).

## IV. EXAMPLE

Transporting things is a main application area for UAVs [26–29]; this includes delivering medicines and immunizations. For the delivery, a transaction message is sent to the UAV with the GPS coordinates and the identifier of the order's package docking device [30]. At the delivery location, the control unit checks to ensure that the identifier matches the one in the transaction message and then performs the package transfer [28–30]. Without loss of generality, we model just the delivery system's pickup, as shown in Fig. 5.

In the figure, a packet-transfer request is created (circle 1) and sent to the UAV system (2), where it is processed (3) to extract the pickup-location address (4). This address—with proper processing (5)—flows (6) to the tracking device (7, the antenna and communication) and is sent (8) to the satellites (9). The pickup location's GPS coordinates then flow from the satellites (10), through the tracking device, to the UAV (11), which processes them (12). This process triggers the creation of control instructions (13). These instructions flow to the actuator (14), which is responsible for moving and controlling the UAV's mechanism. Accordingly, the UAV moves to the pickup location (15).

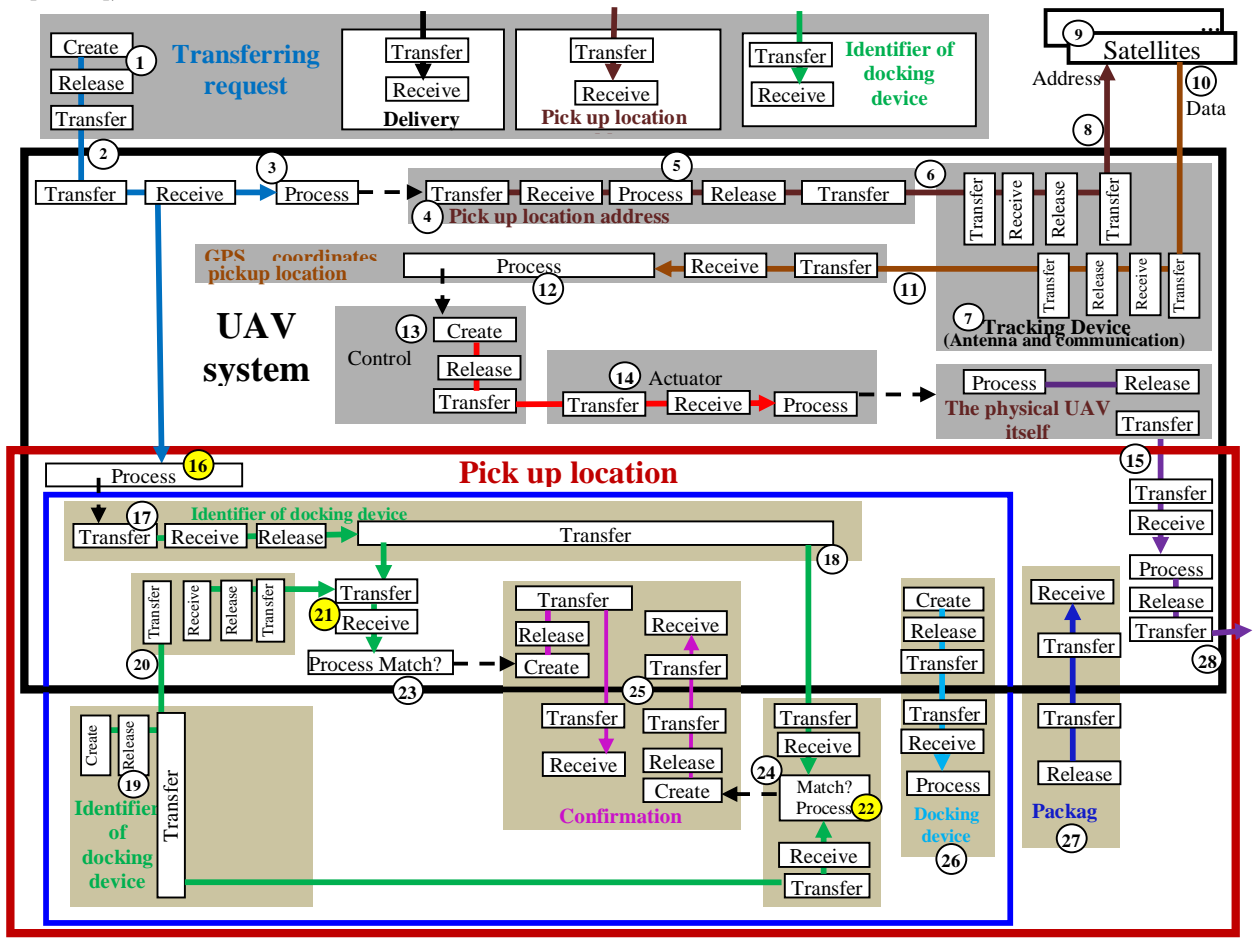

Fig. 5. The TM Model of the UAV Delivery System's Pickup.

Upon arrival at the pickup location, the transfer request is processed again (16) to extract the docking device's identifier (17) and send it to the local location (18). At the same time, the docking device's local-location identifier (19) flows (20) to the UAV, which compares the two identifiers (21, 22). If the identifiers' UAV (23) and local location (24) match, confirmations are exchanged (25).

Upon confirming its identifier from both sides, the docking device (26) moves to the local location and transfers the involved package to the UAV (27). Lastly, the UAV leaves for the delivery location (28).

In a TM model, an event is a machine with at least three submachines: the time, the region, and the event itself. Accordingly, Fig. 6 shows each event, represented by its region, and Fig. 7 shows the UAV system's behavior in terms of the chronology of events (listed below).

- Event$_1$ (E$_1$): A package-transfer request is created.

- Event$_2$ (E$_2$): The request arrives at the UAV, where the local-area address is extracted and sent to the tracking device, which sends the GPS coordinates.

- Event$_3$ (E$_3$): The GPS coordinates are received.

- Event$_4$ (E$_4$): The GPS coordinates are sent to the control, which issues instructions to the actuator.

- Event$_5$ (E$_5$): The UAV is processed (moved) according to the incoming coordinates.

- Event$_6$ (E$_6$): The UAV moves to the pickup location.

- Event$_7$ (E$_7$): The docking-device identifier is extracted from the request and sent to the pickup location.

- Event$_8$ (E$_8$): The docking-device identifier is received and checked at the pickup location.

- Event$_9$ (E$_9$): Confirmations are exchanged between the UAV and the pickup location.

- Event$_{10}$ (E$_{10}$): The package is picked up.

- Event$_{11}$ (E$_{11}$): The UAV reaches the delivery location.

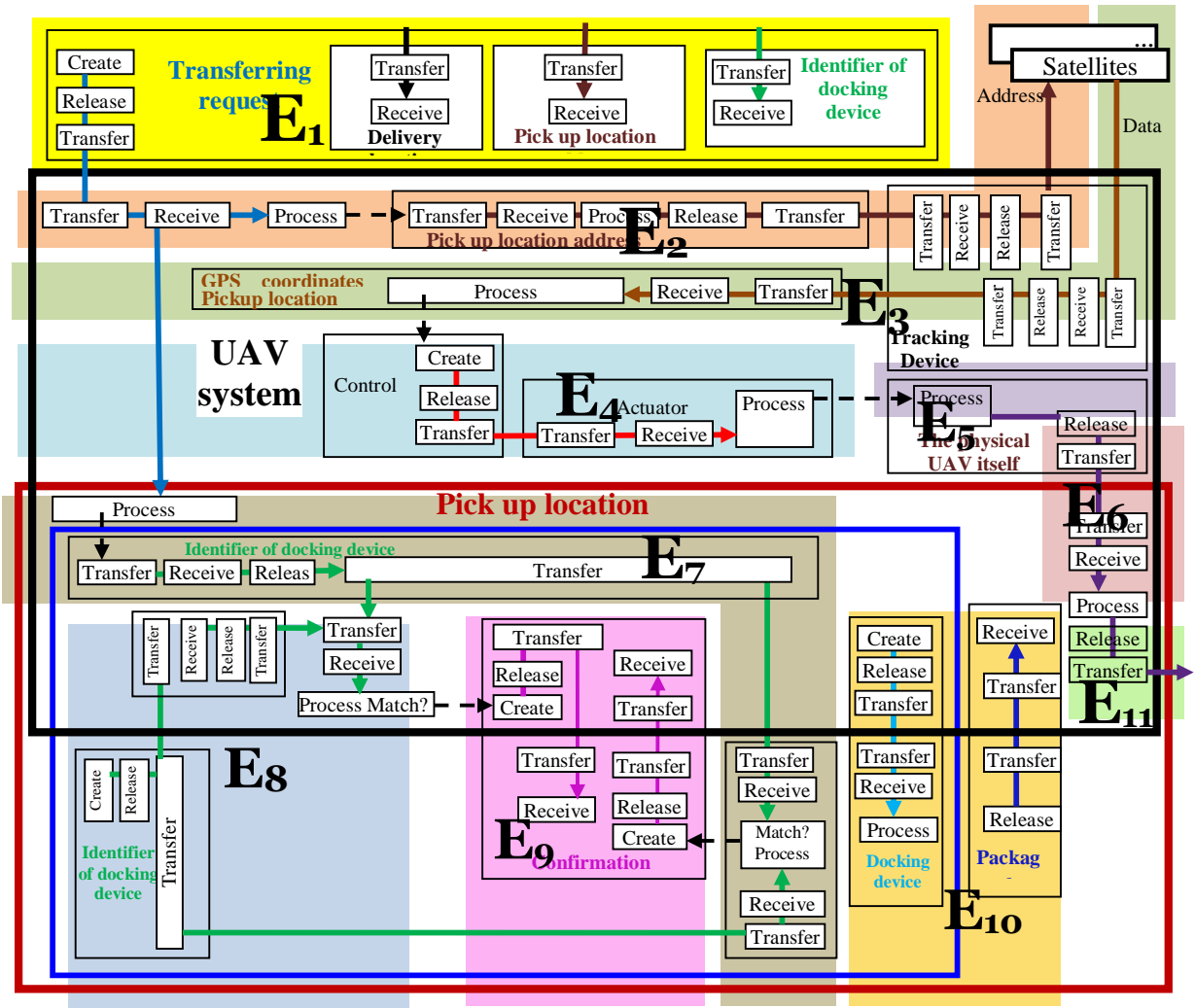

Fig. 6. The TM Model of Part of the UAV Delivery System.

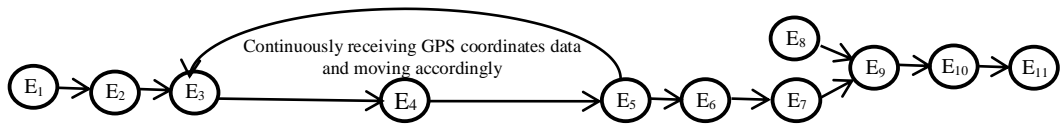

Fig. 7. The Chronology of Events in the TM Model of Part of the UAV Delivery System.

## V. CASE STUDY

This case study is from an actual project called RECON that involves a drone and its control unit, as initially reported in [25]. RECON was implemented to monitor, analyze, inspect, and intervene in data collection for traffic planners, safety managers, and commuters. RECON was originally built using non-TM notions. Fig. 8, 9, and 10 show some of the diagrams used for the original drone (see [25]). No further details about RECON are given because this paper is focused on elaboration of the use of TM and not on description of a UAV project.

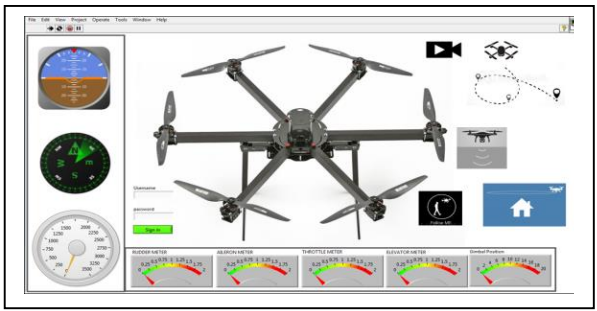

Fig. 8. The RECON user Interface.

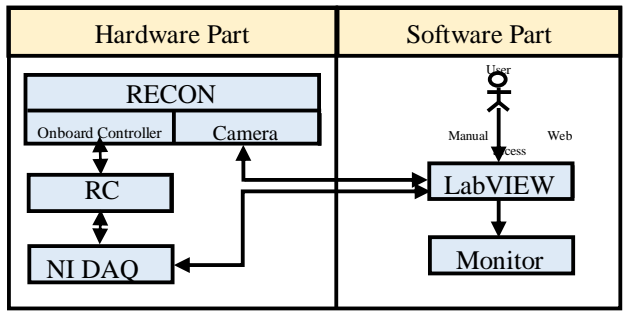

Fig. 9. An Overview of RECON.

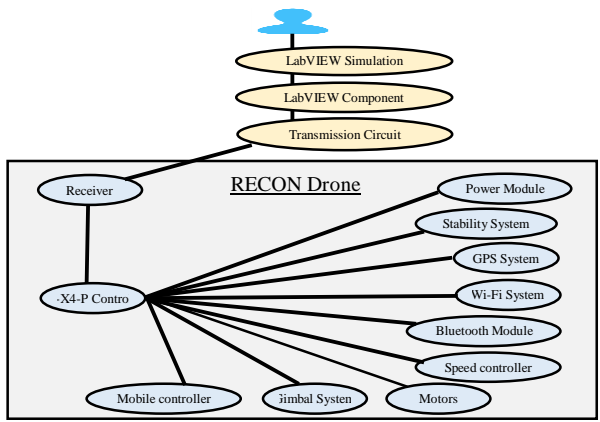

Fig. 10. The RECON user Interface Modeled as a UML use Case.

## VI. MODELING UNMANNED AERIAL VEHICLES

Fig. 11 shows a TM-based static UAV model that consists of a user interface (UI) (circle 1), a server with a control panel (2), and the drone itself—comprising a controller (3) and a physical body (4). The UI contains several pointers (i.e., buttons; see Fig. 8), which are used to manage the flight.

If the drone is turned on, creating a click (5) in the UI causes one of the following signals to be sent (6): point-to-point, auto-landing, elevating signal, lateral balance, throttle signal, rudder signal, fail-safe, and follow-me.

### A. Point-to-Point

This pointer causes the drone to move to a new point (position) from its current point in all directions. To accomplish this point-to-point movement, the UI process (6) creates a signal that is released and transferred (7) to the server, where it is stored (8) in the database. The signal also flows to the drone controller (9), where it is processed to trigger the physical drone's movement (10).

The point-to-point movement (11, in the lower right corner of the figure) creates digital data (12) that flow to the server (13), where they are stored. The data also flow to the UI (14), where they are displayed on the user's screen (15).

### B. Auto-Landing

This pointer causes the drone to land automatically in a given location. To accomplish this, a signal is created, released, and transferred (16) to the server, where it is stored (17). The signal is also sent to the controller (18), where it is processed, which triggers the physical drone's auto-landing operation (19). In addition, the auto-landing operation (20) creates related data (21) that flow to the server (22), where they are stored. The data then flow to the UI (23) for display on the user's screen (24).

### C. Elevating Signal

This pointer causes the drone to move up or down. To accomplish this, a signal is created, released, and transferred (25) to the server, where it is stored (26). The signal then flows to the drone controller (27). In the controller, it is processed to trigger the physical drone's movement (28). Furthermore, this movement (29) creates related data (30) that flow to the server (31), where they are stored. The data then flow to the UI (32), where they are displayed on the user's screen (33).

### D. Lateral Balancing

This pointer measures the rate of rotation and helps keep the drone balanced. Flying with unbalanced props can harm the drone's motors, reduce its flight quality, and affect its stability. Stabilization technology provides navigational information to the controller to enhance flight safety. The drone's lateral balance needs to work almost instantly to act against gravity, wind, and so on. To accomplish this, a signal is created, released, and transferred (34) to the server, where it is stored (35). The signal then flows to the drone controller (36). In the controller, it is processed to balance the physical drone (37).

This stabilizing motion (38) creates related data (39) that flow to the server (40), where they are stored. The data then flow to the UI (41), where they are shown on the user's screen (42).

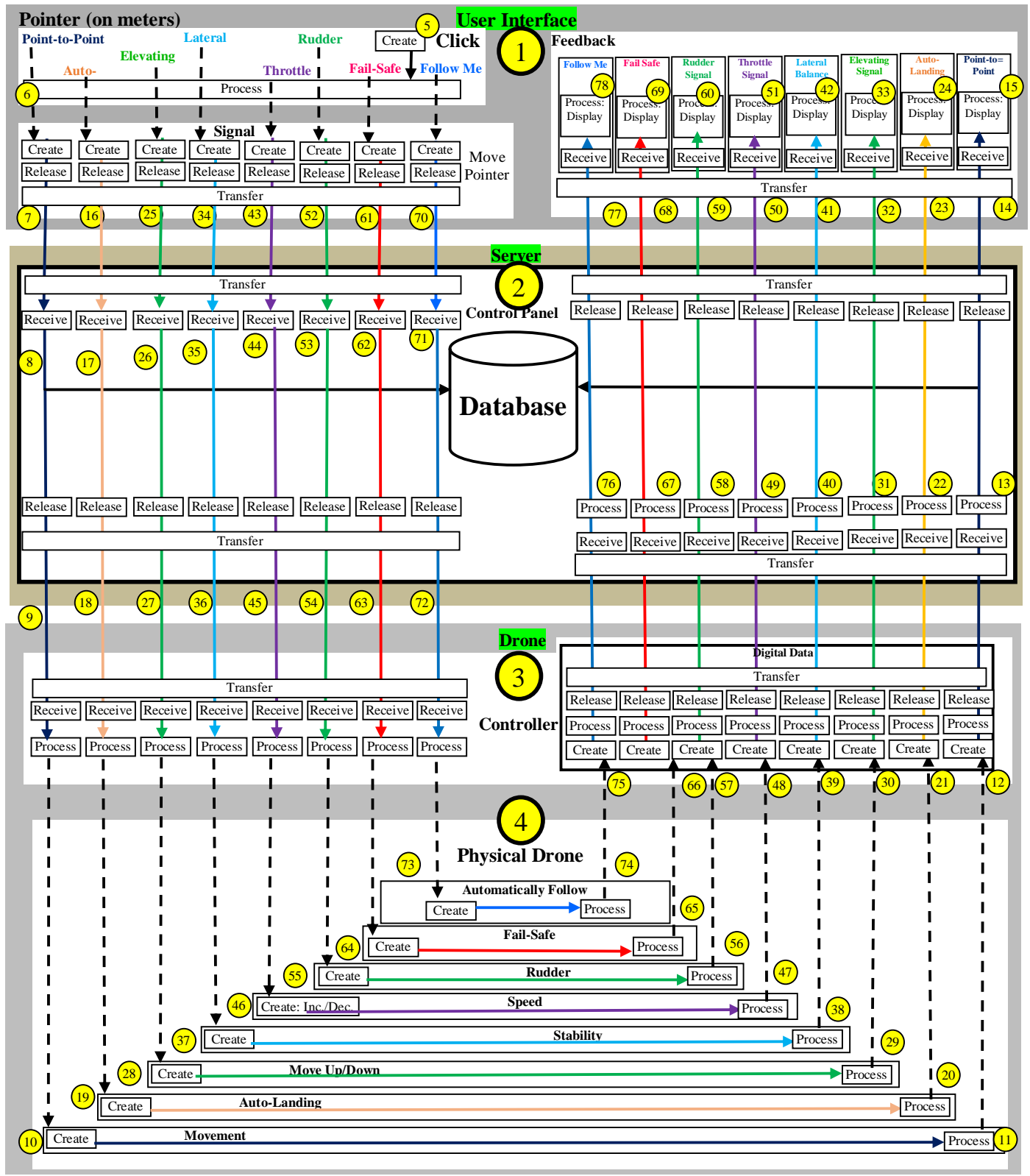

Fig. 11. The TM's Static UAV Model.

*E. Throttle Signal*

This pointer affects the speed of the drone's electric motors (which increase and decrease its speed). Increasing the throttle generates more thrust. To accomplish this, a signal is created, released, and transferred (43) to the server, where it is stored (44). The signal then flows to the drone controller (45), where it is processed to increase or decrease the physical drone's speed (46). In addition, the change in movement speed (47) creates related data (48) that flow to the server (49), where they are stored. The data then flow to the UI (50), where they are displayed on the user's screen (51).

*F. Rudder Signal*

This pointer involves altitude adjustments. To accomplish this, a signal is created, released, and transferred (52) to the server, where it is stored (53). The signal then flows to the drone controller (54), where it is processed to increase or decrease the physical drone's speed (55). In addition, the change in altitude (56) creates related data (57) that flow to the server (58), where they are stored. The data then flow to the UI (59), where they are presented on the user's screen (60).

*G. Fail-Safe*

This procedure supports the drone in case of an error. This mode sets the conditions that the model's servos and motors revert to when it loses the transmitter's control signal. For instance, the fail-safe could automatically cause the drone to return home or to the nearest base station. To accomplish this task, a signal is created, released, and transferred (61) to the server, where it is stored (62). The signal then flows to the drone controller (63), where it is processed to move the physical drone (64). In addition, this movement (65) creates related data (66) that flow to the server (67), where they are stored. The data then flow to the UI (68), where they are displayed on the user's screen (69).

*H. Follow Me*

This pointer causes the activation of follow-me mode, which gives the drone the ability to autonomously track a target without piloting. For example, a drone can be programmed to automatically follow its operator around. To accomplish this, a signal is created, released, and transferred (70) to the server, where it is stored (71). The signal then flows to the drone controller (72), where it is processed to move the physical drone in follow-me mode (73). In addition, this movement (74) creates related data (75) that flow to the server

(76), where they are stored. The data then flow to the UI (77), where they are shown on the user's screen (78).

## VII. DYNAMIC THINGING MACHINE MODEL

The space limitations and density of overlapping events do not permit us to diagram all the events for the processes in the UI. Accordingly, we show the events for only two of them: Point-to-Point and Follow Me, as shown in Fig. 12. Clicking on Point-to-Point causes the following events.

- Event 1 ($E_1$): *Point-to-Point* is clicked on the pointer.

- Event 2 ($E_2$): A flow signal to the server is created.

- Event 3 ($E_3$): The signal is stored in the database.

- Event 4 ($E_4$): The signal flows to the drone controller.

- Event 5 ($E_5$): The physical drone begins to move.

- Event 6 ($E_6$): The movement operation takes its course.

- Event 7 ($E_7$): The signal is processed to trigger up or down movement in the physical drone.

- Event 8 ($E_8$): The signal is processed to balance (stabilize) the physical drone.

- Event 9 ($V_9$): The signal is processed to accelerate or slow the physical drone's movement.

- Event 10 ($E_{10}$): The signal is processed to increase or decrease the physical drone's speed.

- Event 11 ($E_{11}$): The signal is processed to trigger the physical drone's movement; instructions are sent to the flight controller to execute to this mode.

- Event 12 ($E_{12}$): The signal is processed to move the physical drone and thus follow the selected target.

- Event 13 ($E_{13}$): The signal is processed to trigger flows to the controller and shift from analog to digital.

- Event 14 ($E_{14}$): The signal flows to the control panel.

- Event 15 ($E_{15}$): The signal is stored in the database.

- Event 16 ($E_{16}$): The signal flows to the server.

- Event 17 ($E_{17}$): The signal flows to the UI.

Fig. 13 shows the chronology of these events.

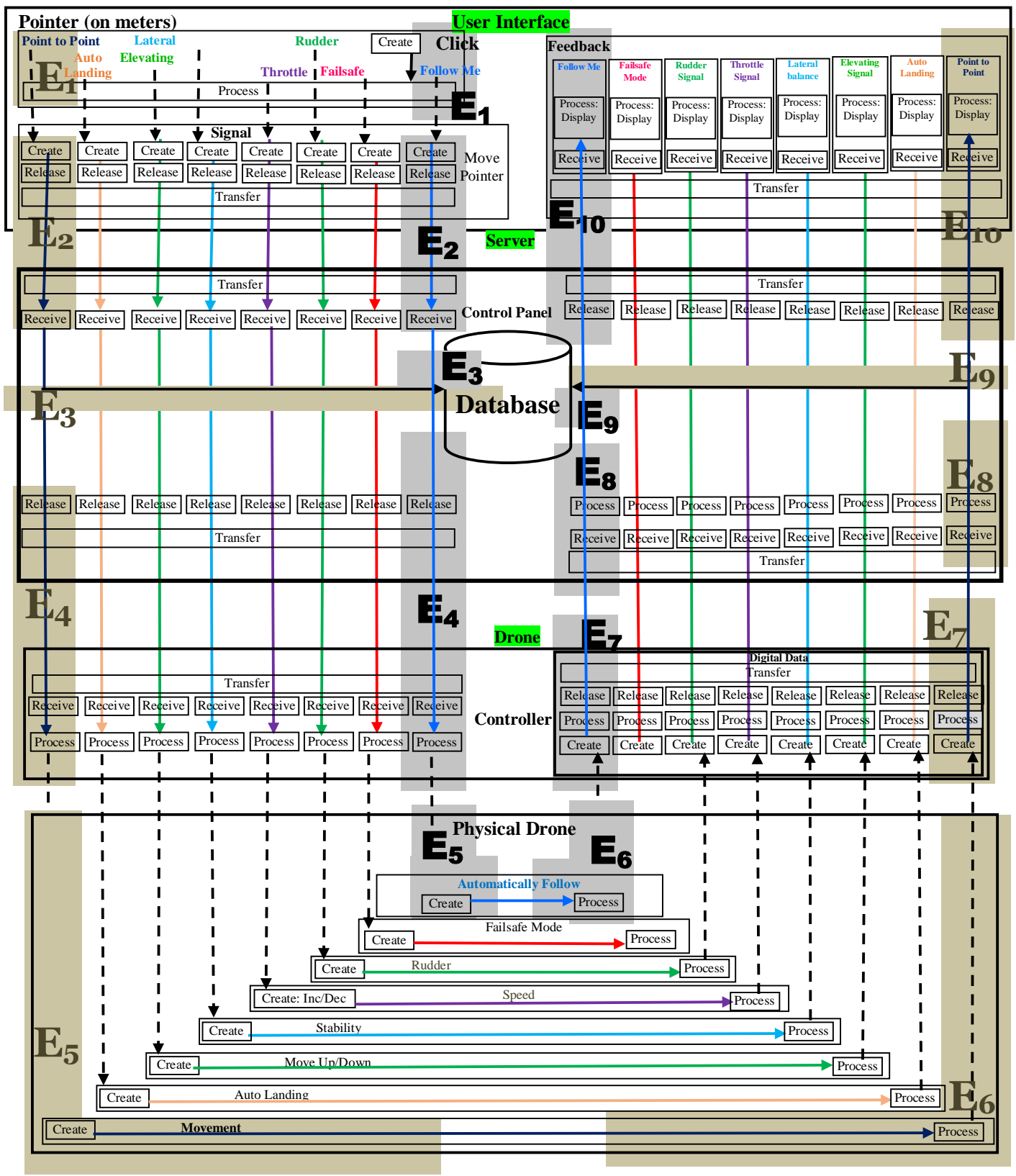

Fig. 12. Events in the TM Model of the UAV Delivery System.

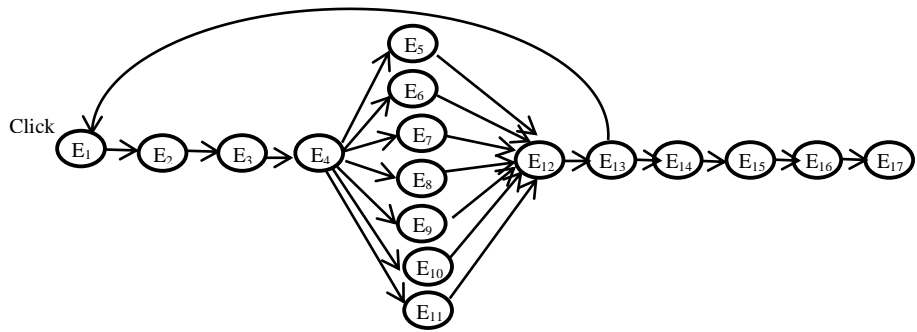

Fig. 13. The Chronology of Events in the TM Model of the UAV.

## VIII. SIMULATION

The TM diagram can be used to simulate drone processes. TM events are fine-grained activities that result in the integration of a static description and a dynamic model of events. Without loss of generality, we focus on flowcharting using the simulation language Arena. In Arena, the flowchart plays an important role, and the success of a simulation depends on how well the flowchart projects the identification of events; the notion of an event is used informally. TM can assist with this process.

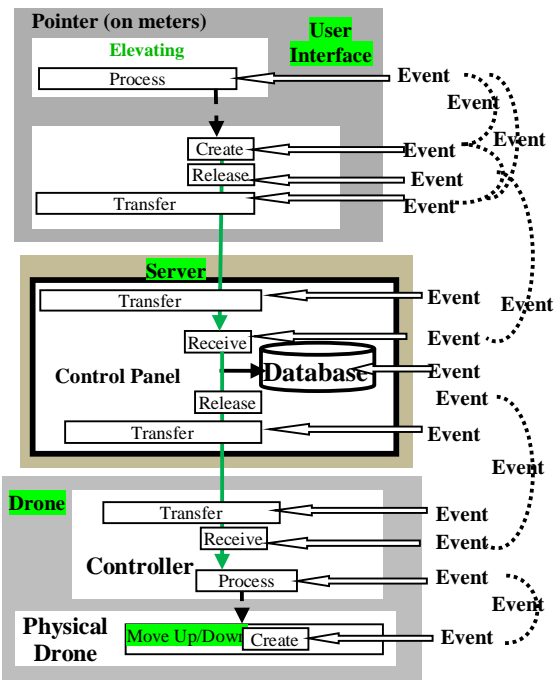

Fig. 14. Some Combinations in a Series of Elementary Elevating Events.

By contrast, a TM diagram specifies all elementary events (create, process, release, transfer, and receive). For example, Fig. 14 shows the series of elementary events involved in elevating the drone (see Fig. 11, circle 5), including the click and the movement of the drone, as well as some more complex events that can be formed from the elementary events. Many possible selections of events are possible. For example, release-transfer/transfer-receive can be considered one event in which a thing flows from one submachine to another. Alternatively, it can be considered two events: leaving (release-transfer) and arriving (transfer-receive). We start with a TM diagram and identify event boundaries from the elementary events until we attain the required level of granularity. We are experimenting with Arena flowcharts produced by an ad hoc method as well as those developed using TM.

## IX. CONCLUSION

This paper addresses the development of conceptual modeling for UAVs. We propose a flow-based specification called TM as a good vehicle in this area and demonstrate the TM methodology through a case study involving the construction of a drone.

A shortcoming of TM regards its (visual) diagramming complexity, which originated in the various machines' and submachines' completeness. The TM diagram can be simplified to whatever level of granularity is required for the original TM description. For example, Fig. 15 was produced from a static TM representation of a UAV.

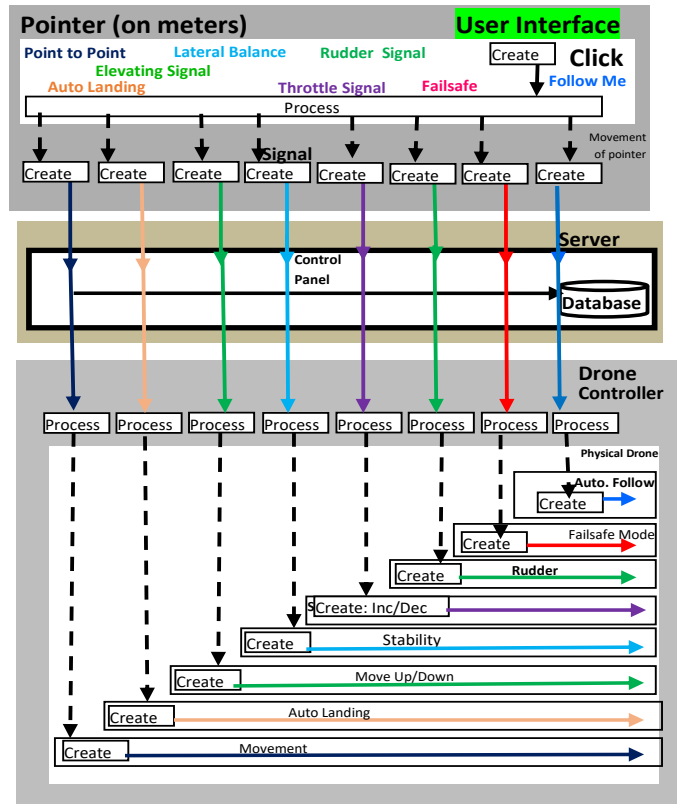

Fig. 15. Simplification of the TM's Static UAV Model (Partial).

Further research is needed to directly apply the TM methodology to more sophisticated UAV systems. Further investigation is also required to develop TM tools and supporting apparatus. Specifically, additional synchronization, constraints, and logical notation need to be superimposed on the base TM description. As mentioned previously, TM can be used as a basis for simulation.